\documentclass{aa}
\usepackage{txfonts}
\usepackage{graphicx}
\usepackage{hyperref}
\def\asec{\ifmmode ^{\prime\prime}\else$^{\prime\prime}$\fi}

\def\it{\sl}
\def\degs{\ifmmode ^{\circ}\else$^{\circ}$\fi}
\def\amin{\ifmmode ^{\prime}\else$^{\prime}$\fi}
\def\asec{\ifmmode ^{\prime\prime}\else$^{\prime\prime}$\fi}
\def\farcs{\hbox{$.\!\!^{\prime\prime}$}}  % Fractions of arcseconds
\def\degs{\ifmmode ^{\circ}\else$^{\circ}$\fi}
\def\amin{\ifmmode ^{\prime}\else$^{\prime}$\fi}

\def\eqalign#1{\null\,\vcenter{\openup1\jot \m@th
   \ialign{\strut\hfil$\displaystyle{##}$&$\displaystyle{{}##}$\hfil
   \crcr#1\crcr}}\,}
\begin{document}

\markboth{Time-resolved observations of  SDSS J123813.73-033933.0 }{Zharikov et al: Time-resolved observations of 
 SDSS J123813.73-033933.0 }

% Some definitions I use in these instructions.
\authorrunning{S. Zharikov, G. Tovmassian, et al.}
\titlerunning{Observations of CV SDSS J123813.73-033933.0}
\title{Time-resolved observations of the short period~CV \\ SDSS J123813.73-033933.0}
\author{S.V. Zharikov\inst{1}
\and G. H. Tovmassian\inst{1}\thanks{Visiting
research fellow at Center for Astrophysics and Space Sciences,
University of California, San Diego, 9500 Gilman Drive,
La Jolla, CA 92093-0424, USA}
\and R.~Napiwotzki\inst{2}
\and R. Michel\inst{1}
\and V. Neustroev\inst{3,4}
}
\institute{
Observatorio Astronomico Nacional SPM, Instituto de Astronomia, UNAM, Ensenada,
BC, Mexico \\
zhar,gag,rmm@astrosen.unam.mx
\and
 Centre for Astrophysics Research, University of Hertfordshire, College
Lane, Hatfield AL10 9AB, UK \\
rn@star.herts.ac.uk
\and
Computational Astrophysics Laboratory, National University
of Ireland, Galway, Newcastle Rd., Galway, Ireland \\       benj@it.nuigalway.ie
\and              Isaac Newton Institute of Chile, Kazan Branch      }

\offprints{S. Zharikov,\\
\email{zhar@astrosen.unam.mx}}

\date{Received --- 9 June 2005, accepted --- 11 Octuber 2005}

\abstract{} {We observed a new and poorly studied cataclysmic variable (CV)
\object{SDSS~J123813.73-033933.0} to  determine its classification and
binary  parameters.  }{  Simultaneous  time-resolved  photometric  and
spectroscopic observations were carried out to conduct period analysis
and Doppler tomography mapping.}   { From radial velocity measurements
of  the  H$\alpha$  line  we   determined  its  orbital  period  to  be
$0.05592\pm0.00035$ days  (80.53min).  This period is  longer than the
first estimate  of 76 min  by Szkody et al.  (2003), but
still at the very edge of the period limit for hydrogen-rich CVs.  The
spectrum shows  double-peaked Balmer emission lines  flanked by strong
broad  Balmer absorption,  indicating  a dominant  contribution by  the
white dwarf primary star, and is similar to the spectra of short-period
low-mass  transfer WZ Sge-like  systems.  The photometric  light curve
shows complex variability.  The system undergoes cyclic brightening up
to  0.4 mags, which  are of  semi-periodic nature  with periods  of the
order  of 8-12  hours.   We also  detect  a 40.25  min variability  of
$\sim0.15$ mag corresponding to half of the orbital period.  Amplitude
of the latter increases with the cyclic brightening of the system.  We
discuss the variable accretion rate and  its impact on the hot spot as
the   most   probable    reason   for   both   observed   processes.}{
\object{SDSS~J123813.73-033933.0}  is   preliminary  classified  as  a
WZ\,Sge-like  short  period system  with  low  and unstable  accretion
rate.}{}{

 } 

\maketitle

\section{Introduction}
Cataclysmic variables  (CVs) are close  binaries that contain  a white
dwarf (WD)  and a  late main-sequence (K-M  spectral type)  star.  The
secondary  star fills  its Roche  lobe and  transfers mass  to  the WD
through  the inner  Lagrangian point.   More than  1300 CVs  are known
presently (Downes et al.  \cite{Downes}) and orbital periods have been
found for more  than 400 systems. The orbital  periods range from days
to a minimum  period of about 70 min for systems  with a main-sequence
secondary star.  An  important feature that stands out  in the orbital
period distribution of  CVs is a sharp short-period  cut-off at 80min,
the 'period minimum' (e.g. Barker \& Kolb \cite{Barker}).  Less than a
dozen AM CVn-type CVs with even shorter periods are interpreted as CVs
with helium star donors.  According to the population syntheses, there
should be  a significant  number of systems  near the  minimum period,
which  have not  been observed  (Kolb \&  Baraffe  \cite{Kolb}).  Some
calculations predict that 99\% of the entire CV population should have
orbital  periods $<2$h (e.g.   Howell et  al. \cite{Howell}),  but the
number of CVs observed above and below the period gap are similar.  If
the population models  are correct, then only a  small fraction of the
existing CV population has been discovered so far.
 
 About 400 new  CVs are expected from the  complete realisation of the
Sloan  Digital Sky  Survey (SDSS).   The  four releases  of SDSS  have
unveiled   132   new   CVs   ({Szkody   et   al.    \cite{Szkody2002},
\cite{Szkody2003}, \cite{Szkody2004}, \cite{Szkody2005})}.  Among them
is  \object{SDSS  J123813.73-033933.0},  which  was identified  as  an
r=17.82  magnitude CV  (u=17.89, g=17.78,  i=17.97, z=18.07)  with the
extremely short orbital period of 76 min determined from seven spectra
obtained  by   Szkody  et  al.    (\cite{Szkody2003}).   The  galactic
coordinates of the object are $l=296.51$ and $b=+59.05$, which implies
a  galactic extinction  factor of  only E(B-V)=0.03  (Schlegel  et al.
\cite{Schlegel}).    The   proper    motion   was   measured   to   be
$p.m.=0\farcs143$/year as presented in  the USNO B1.0 catalogue (Monet
et al.  \cite{Monet}).  The  optical spectrum of  this system  shows a
blue  continuum with  broad absorption  features  around double-peaked
Balmer emission  lines.  The spectral appearance  resembles the small,
but intriguing, group of so-called \object{WZ\,Sge} objects remarkable
for  their  large amplitude  outbursts,  long  recurrence cycles,  and
peculiar  outburst lightcurves.  {They are  concentrated close  to the
lower limit  of the  CV period distribution  and are believed  to have
bounced back from the minimum-period limit; i.e. dwarf novae that have
periods  that  are  lengthening  after  evolving  through  the  period
minimum.}  These  are supposed to  be those very missing  systems with
short  periods  that, according  to  theoretical calculations,  should
produce a spike  in the number of CVs at  the period minimum. However,
the  discovery  and study  of  them  is  hindered by  their  intrinsic
faintness,  due  to  the   low-mass  transfer  rates,  and  infrequent
outbursts.

  The  subject   of  this   paper  is  a   time-resolved  simultaneous
spectroscopic   and   photometric  study   of   the  CV   \object{SDSS
J123813.73-033933.0} (hereafter abbreviated as \object{SDSS1238}).  In
Sect.\ref{Obs} we  describe our  observations and the  data reduction.
The data  analysis and the  results are presented  in Sect.\ref{DatAn}
while a discussion  and a summary are given  in Sects.\ref{Discus} and
\ref{Summ}, respectively.
 
\begin{table*}[]
\begin{center}
\caption{Log of observations of SDSS J123813.73-033933.0.}
\begin{tabular}{llllccc}
\hline\hline
 Date (2004y.)       & HJD Start+ & Telescope& Instrument/Grating  &Range/Band & Exp.Time/Num. of Integrations& Duration\\ 
Spectroscopy &  2453000       &   &        &     &     &        \\ 
14 Apr & 109.868 &2.1m &B\&Ch$^1$ 1200l/mm  & 6000-7100\AA    &  600s$\times$11   & 1.92h  \\
15 Apr & 110.650 &2.1m &B\&Ch\ \ \ 1200l/mm  & 6000-7100\AA    &  600s$\times$22   & 3.84h  \\
16 Apr & 111.724 &2.1m &B\&Ch 400l/mm  &  4200-7300\AA   & 600s$\times$15    & 2.93h  \\ 
17 Apr & 112.658 &2.1m &B\&Ch 400l/mm   & 5200-8200\AA    &  600s$\times$20  & 3.27h  \\
17 Apr & 112.852 &2.1m& B\&Ch 400l/mm   & 3700-6800\AA   &   900s$\times$11  & 4.65h  \\
19 May & 145.710 &2.1m &B\&Ch 400l/mm   &  4200-7300\AA    &   600s$\times$23  & 4.32h  \\
Photometry &  &   &    & &                                               &                \\
15 Apr  & 110.737 &1.5m & RUCA$^2$  & R    & 180s$\times$101                     & 5.76h  \\
16 Apr & 111.751  &1.5m & RUCA & R    &  180s$\times$119                    & 5.18h  \\ 
17 Apr & 112.653  &1.5m & RUCA & V    & 120s$\times$129                        & 7.22h  \\
16 May & 142.675  &1.5m & RUCA & V    & 120s$\times$133                     & 5.38h  \\
17 May & 143.663  &1.5m & RUCA & V    & 120s$\times$128                     & 5.51h  \\      
18 May & 144.646  &1.5m & RUCA & V    & 120s$\times$143                     & 5.80h  \\
19 May & 145.652  &1.5m & RUCA & V    &120s$\times$135                      & 5.64h  \\ \hline
\end{tabular}
\label{tab1}
\end{center}
\begin{tabular}{l}
$^1$ B\&Ch - Boller \& Chivens spectrograph (http://haro.astrospp.unam.mx/Instruments/bchivens/bchivens.htm) \\
$^2$ RUCA - CCD photometer (http://haro.astrospp.unam.mx/Instruments/laruca/laruca\_intro.htm)
\end{tabular}
\end{table*}

\section{Observations}
\label{Obs}
The  observations  of  SDSS1238  were  obtained  at  the  Observatorio
Astronomico  Nacional  (OAN SPM\footnote{http://www.astrosen.unam.mx})
in  Mexico.   Photometric  observations  were performed  on  the  1.5m
telescope covering a longer time span:  3 nights in April and 4 nights
in  May  2004.  Differential  photometry  using  four  field stars  as
reference  was  obtained in  the  V  and  R broadband  Johnson-Cousins
filters in long time series with individual exposures ranging from 120
to 180s.

 The  2.1m telescope  with  the B\&Ch  spectrograph,  equipped with  a
24~$\mu$m ($1024\times1024$)  SITe CCD was used  for the spectroscopic
observations. Spectra were obtained in the first order of 1200 and 400
line/mm  gratings.   A  relatively  narrow  slit was  used  (width  of
1\farcs5) resulting in a spectral  resolution of FWHM 2.3\AA \ and 6.2
\AA  \ for  each corresponding grating.  Spectroscopic observations
were done  during four consecutive nights  in April and one  night in May,
2004  with a total  coverage of  about 21h  with 10  minute individual
exposures. The  He-Ar lamp was taken  every 2-3 hours  during the runs
for wavelength calibrations, and the standard spectrophotometric stars
\object{HZ44}, \object{Feige34}, \object{Feige67}, \object{BD+33 2642}
(Oke  \cite{Oke})   were  observed  (a  pair  each   night)  for  flux
calibrations. At  the end, a total  of 102 spectra  were obtained. The
reduction of  the images and the  extraction of the  spectra were done
using standard IRAF
\footnote{IRAF is the Image Reduction and Analysis Facility, a general
purpose software system for the reduction and analysis of astronomical
data. IRAF is  written and supported by the  IRAF programming group at
the  National  Optical   Astronomy  Observatories  (NOAO)  in  Tucson,
Arizona.  NOAO is  operated  by the  Association  of Universities  for
Research in  Astronomy (AURA), Inc., under  cooperative agreement with
the National Science Foundation}  routines commonly used for long-slit
spectra.   The  bias  subtraction  was applied  using  overscan  strip
correction, and spectra were  extracted using optimal weights based on
the  spatial  profile.   The  wavelength calibration  was  done  using
multiple  arc lamps  taken during  the night  and  flux calibration
using   at   least   two   standard  stars   observed   at   different
airmasses.  During  some  of  these nights  we  obtained  simultaneous
photometry of the object in  the V band on the accompanying telescope.
For  those nights  we corrected  spectra to  corresponding photometric
fluxes. This, however, does  not warrant excellent flux calibration in
the blue part of the spectra, since a narrow slit was used with seeing
that  was usually comparable  to the  slit width  and a  permanent E-W
orientation  of the  slit.   A  log of  observations  is presented  in
Table\,\ref{tab1}.

 \begin{figure*}[t]
 \setlength{\unitlength}{1mm}
 \resizebox{12.cm}{!}{
 \begin{picture}(90,50)(0,0)
 \put (-3,0){\includegraphics[width=125mm,bb=30 50 675 320, clip=]{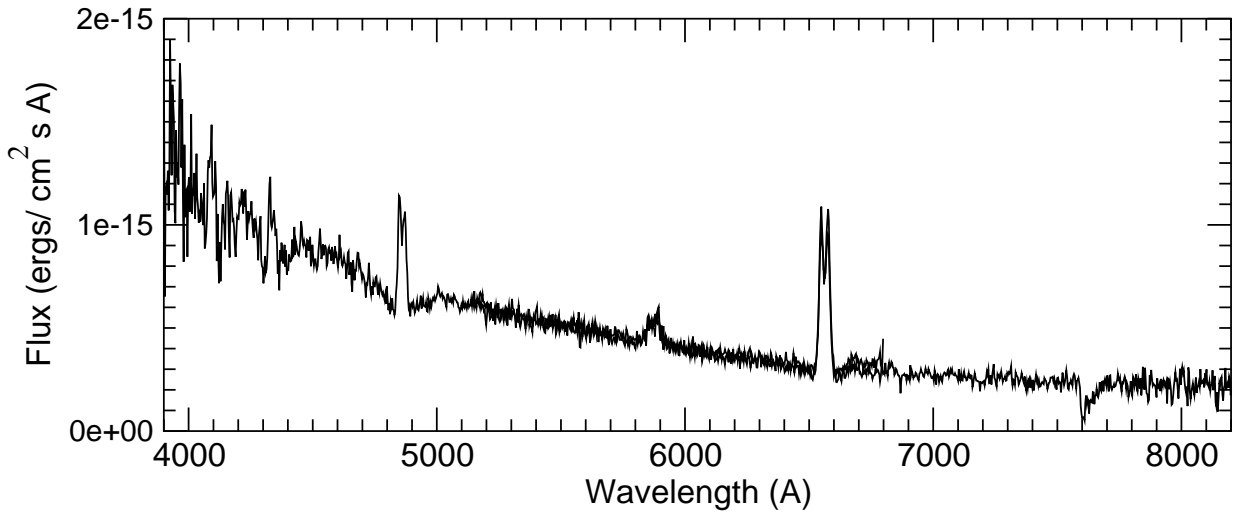}}
 \end{picture}}
 \caption{The low-resolution time-averaged composite spectrum of SDSS1238 
 (10 spectra $3900-6800$ \AA \  in 
 blue and 20 spectra $5200-8200$ \AA \ in red ranges corrected for 
 radial velocity shifts due to orbital motion) is shown.  
 }
 \label{fig1}
 \end{figure*}

\section{ Data analysis.}
\label{DatAn}

\subsection{Spectrum of SDSS1238}
 The  time-averaged  spectrum of  \object{SDSS1238}  corrected for  K1
 velocity shifts in the white dwarf frame due to the orbital motion is
 shown  in Fig.\,\ref{fig1}.   The spectrum  shows  the characteristic
 features  of a  short-period, low-mass  transfer rate  and relatively
 high  inclination  cataclysmic variables,  like  \object{WZ Sge}  and
 \object{RZ  Leo}.   The continuum  of  the \object{SDSS1238}  roughly
 resembles the spectrum of a DA3 type WD.  There is no evidence of the
 presence of the secondary star in the red part of the spectrum, which
 is also common  in this kind of object.  There  are strong Balmer and
 weak HeI,  HeII emission lines.   All emission lines in  the spectrum
 have wide  double-peaked profiles, suggesting that  they are produced
 in  a  high inclination  accretion  disc  and/or  the combination  of
 emission lines from  the disc and the underlying  absorption from the
 white  dwarf   primary  (see   Fig.\,   \ref{fig1}).  The  symmetric
 absorption at the wings of  emission lines in addition to the central
 dip  indicate  a  rather   stellar  origin  of  the  absorption.  The
 double-peaked  profile structure  of  the Balmer  lines is  preserved
 during all orbital phases of the system.

{The shape of  the broad absorption lines agrees  with their formation
in  the high  density photosphere  of  the WD.   Similar broad  Balmer
absorption lines have been detected  in a number of other short period
dwarf novae (e.g.  WZ Sge, GW Lib,  BC UMa, or BW Scl).   We could not
completely  rule out  a geometrically  thin but  optically  thick disc
origin   of   observed    absorptions;   however,   the   hypothetical
identification of the observed  Balmer absorptions as the photospheric
spectrum  of the  white dwarf  has been  confirmed by  the unambiguous
detection of the white  dwarf at ultraviolet wavelengths (G\"{a}nsicke
et  al.  \cite{Gans},  Szkody   et  al.  \cite{Szkody2002b},  Sion  et
al. \cite{Sion90}) in all above mentioned systems. } We determined the
WD parameters by fitting Balmer  lines of the observed spectrum with a
grid of  a pure hydrogen model  spectra calculated with  the NLTE code
developed  by Werner  (\cite{Werner}).   The model  grid  and the  fit
procedure are described in Napiwotzki (\cite{Napiwotzki}).  Deviations
from LTE are unimportant in the parameter range explored here, but the
model atmospheres contain the  input physics required for the analysis
of DA white dwarfs.

\begin{figure}[t]
\setlength{\unitlength}{1mm}
\resizebox{8cm}{!}{
\begin{picture}(100,115)(0,0)
\put (10,0){\includegraphics[width=90mm,bb =20 20 590 750, clip=]{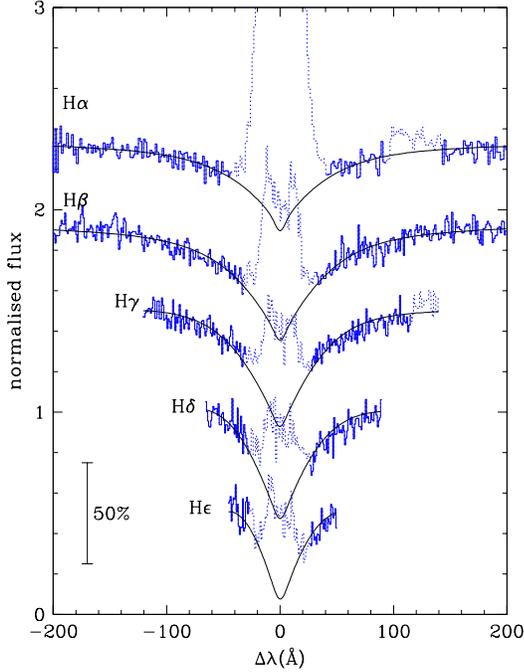}}
\end{picture}}
\caption{ The WD atmosphere model fitted to the Balmer absorption lines of  SDSS1238.}
\label{fig2}
\end{figure}

Since the signal-to-noise of our co-added spectrum in the blue/UV part
was   low,  we  performed   the  spectral   analysis  with   the  Sloan
Survey\footnote{ Funding for the creation and distribution of the SDSS
Archive  has been  provided by  the  Alfred P.  Sloan Foundation,  the
Participating  Institutions,   the  National  Aeronautics   and  Space
Administration, the National  Science Foundation, the U.S.  Department
of Energy,  the Japanese Monbukagakusho,  and the Max  Planck Society.
The SDSS  Web site  is http://www.sdss.org/.}  spectrum.   The regions
contaminated  by emission  from the  disc were  excluded from  the fit
(cf.\  Fig.\,\ref{fig2}).   Due  to   the  disc  contamination  it  is
difficult to  determine temperature and  gravity simultaneously.  Thus
we preferred to fix gravity at a range of values (allowed range: $\log
g  =7.6...8.2$)  and only  fit  the  temperature.   The combined  best
estimate  of the  WD parameters  is $T_{\mathrm{eff}}=15600\pm1000$\,K
and $\log g = 7.85\pm0.25$.

Note that cooler solutions with  temperatures below the maximum of the
Balmer lines  exist as well ($T_{\mathrm{eff}}$ in  the range 10000\,K
to 11000\,K). The  fits are slightly worse, but  the difference in fit
quality is not large enough to rule out the cool solutions. Additional
information  is provided by  the photometric  colours observed  in the
course  of the  Sloan survey:  $u-g =  0.06$ and  $g -r  =-0.05$.  The
relations of Smith et  al. (\cite{Smith1}) allow a transformation into
Johnson   colours:   $U-B=-0.79$    and   $B-V=0.15$.    Bergeron   et
al. (\cite{Bergeron}) computed colours  for white dwarfs. A comparison
of  the observed  colours of  \object{SDSS1238} with  their tabulation
yields  $\approx$ 17000\,K  from the  $U-B$ colour  and  13000\,K from
$B-V$. Both  results are  only consistent with  our hot  solution. The
large  difference  between  the  temperatures computed  from  the  two
colours are  probably explained by  the contribution of the  disc. The
contribution of  the Balmer emission  lines increases the flux  in the
$g$ and $r$  filters.  For a simple estimate we  produced a version of
the observed  spectrum with the  emission lines removed by  hand.  The
spectra  were  convolved  with   the  transmission  curves  (Smith  et
al. \cite{Smith1})  and magnitudes computed  for the spectra  with and
without emission lines. The result is that the emission lines increase
the  flux in  $g$ and  $r$ by  approximately 0.03\,mag  and 0.06\,mag,
respectively. This causes a  corresponding reddening of the colours of
the  system as  compared to  an isolated  white dwarf.   An additional
contribution  comes  from the  continuum  of  the  disc; however,  the
current observational data does not allow us to perform a quantitative
estimate.

\begin{figure*}[t]
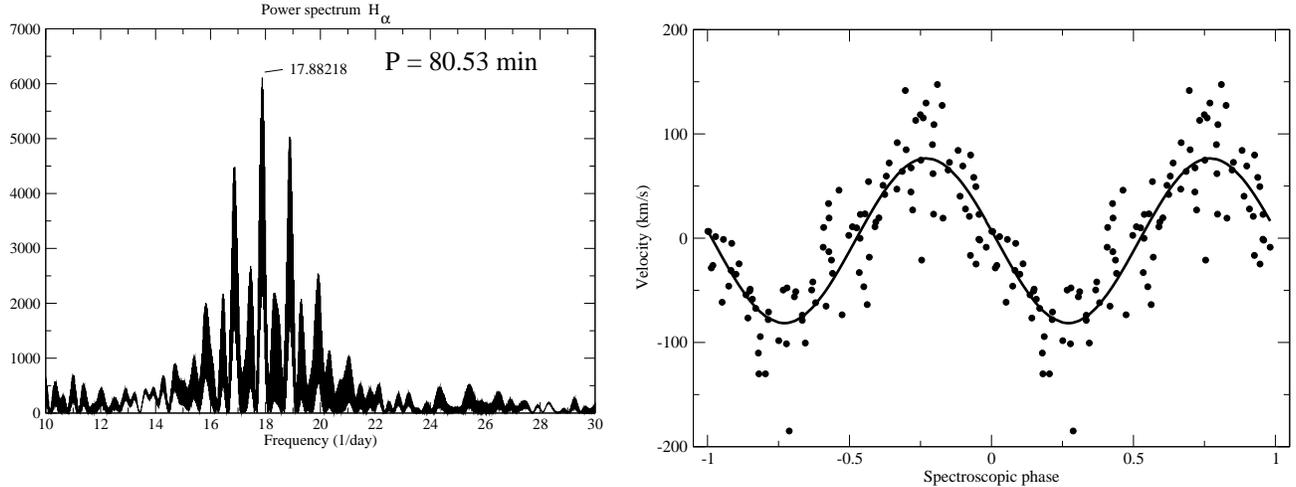

\vspace*{0.2cm}
\setlength{\unitlength}{1mm}
\resizebox{8cm}{!}{
\begin{picture}(100,80)(0,0)
\put (10,0){\includegraphics[width=100mm,bb=50 10 715 550, clip=]{zharikovfig3a.eps}}
\put (115,0){\includegraphics[width=110mm, clip=]{zharikovfig3b.eps}}
\end{picture}}
\caption{{ Left)} The power spectrum of the H$\alpha$ radial velocity curve. The maximum peak in frequency 
corresponds to the orbital period of the system P$_{\rm orb}$=0.05592 day.
{ Right)} The radial velocity measurements of the  H$\alpha$ emission line 
 folded with the spectroscopic
 orbital period (0.05592 days) and its best fit sinusoid.}
\label{fig3}
\end{figure*} 

\begin{figure*}[t]
\vspace*{0.2cm}
\setlength{\unitlength}{1mm}
\resizebox{14.2cm}{!}{
\begin{picture}(100,130)(0,0)
\put (20,0){\includegraphics[width=90mm,bb=50 10 580 780, clip=]{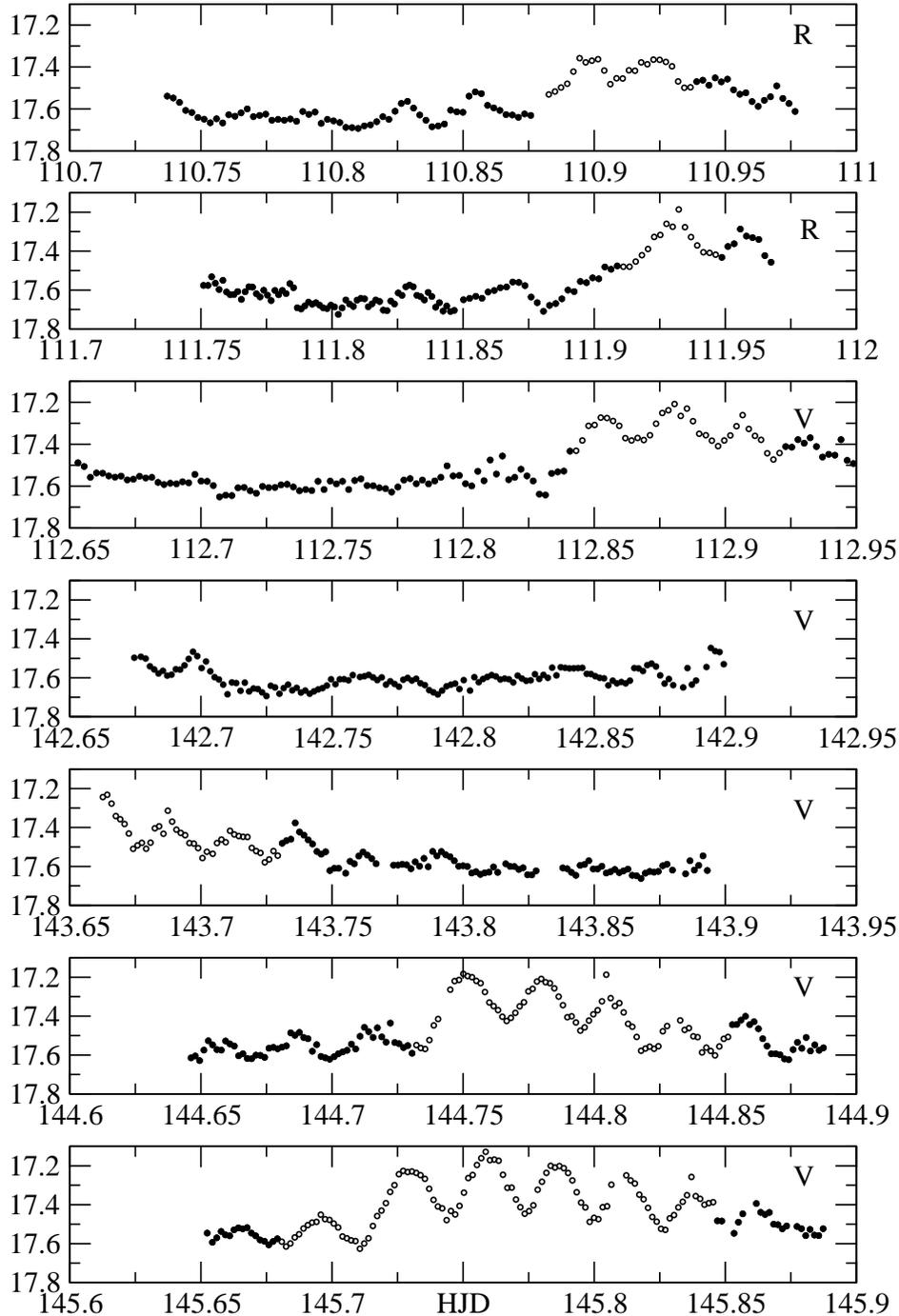}}
\end{picture}}
\caption{ The   light curves of SDSS1238  in $V$ and $R$ bands. The reference  for magnitude is
the USNO B1 value of SDSS1238 r=17.6  arbitrarily assigned to the  minimum of the
measured magnitudes.  Each night is presented in a separate panel. The open circles mark 
 the data used for the period search of 
short-term variability (see text).}
\label{fig4}
\end{figure*} 
\begin{figure*}[t]
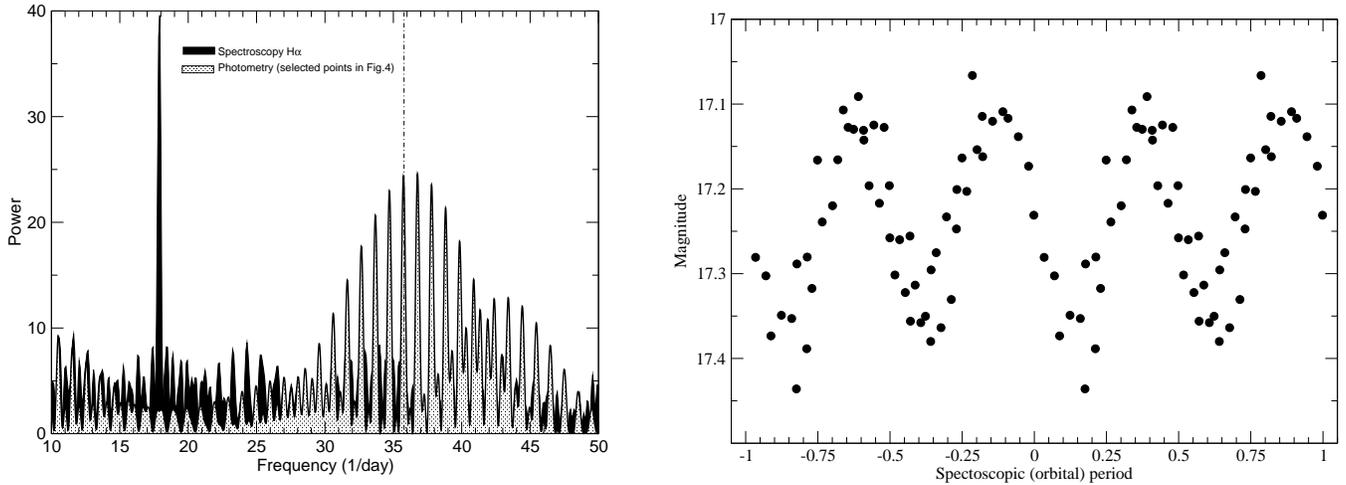

\setlength{\unitlength}{1mm}
\resizebox{8cm}{!}{
\begin{picture}(100,85)(0,0)
\put (0,0){\includegraphics[width=110mm,bb=50 10 715 550, clip=]{zharikovfig5a.eps}}
\put (115,5){\includegraphics[width=112mm,bb=30 50 705 560, clip=]{zharikovfig5b.eps}}
\end{picture}}
\caption{ Left) The CLEANed power spectrum of $H\alpha$ RVs and power spectrum  of  selected photometry points  from the bright
part of the light curve (see Fig.\,\ref{fig4}) . 
 The photometric period  is 2 times
shorter than the orbital period.  Right) The  light curve comprised of the selected points 
(19 May, HJD between 2453145.75 and 2453145.84 with long-term variability trend removed)
folded with the short-term
variability period. The maximum semi-amplitude of the
variability is about 0.2 mag.}
\label{fig5}
\end{figure*} 
\begin{figure*}[t]
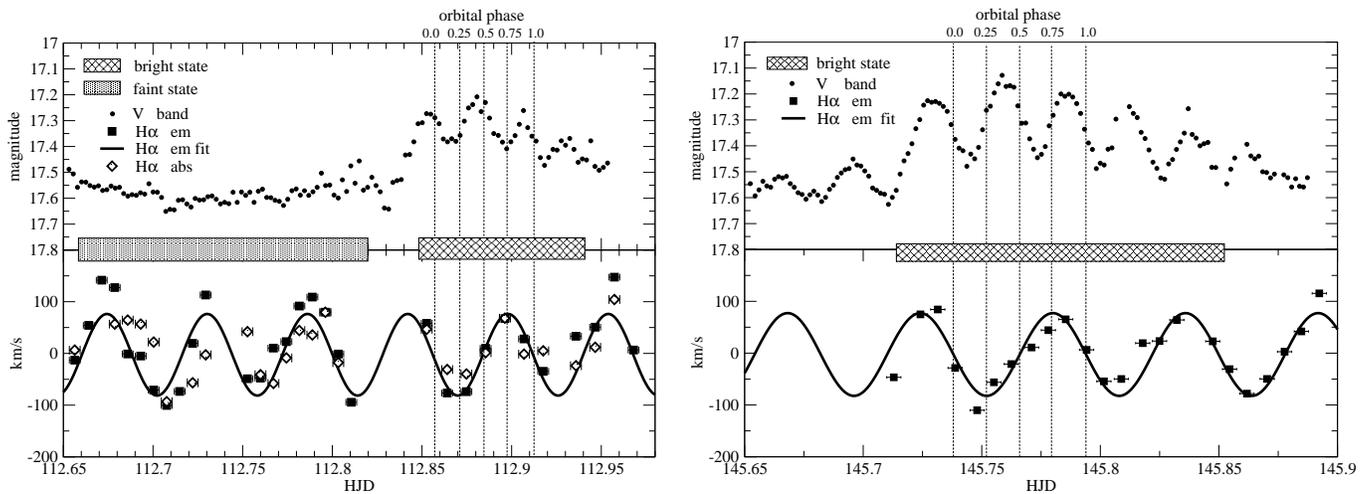

\setlength{\unitlength}{1mm}
\resizebox{8cm}{!}{
\begin{picture}(100,80)(0,0)
\put (0,0){\includegraphics[width=110mm, clip=]{zharikovfig6a.eps}}
\put (115,0){\includegraphics[width=113mm, clip=]{zharikovfig6b.eps}}
\end{picture}
}
\caption{The lights curves (top panels) and RV of  H$\alpha$ with corresponding best fit
curves (low panels) of SDSS1238 for 17 April (right) and 19 May
(left). The full squares (low panels) and open diamonds  (low right panel)
present the H$\alpha$ emission and absorption(central)  components, respectively. Vertical dashed
lines show selected orbital phases of the system. Two shaded bands mark  portions  of  data
in the bright and faint states of the SDSS1238 light curves used for
Doppler tomography analysis.}
\label{fig6}
\end{figure*}

\subsection{Orbital period determination.} 

 In order to  determine the orbital parameters, we  measured the radial
velocities  (RV)  from  H$\alpha$   and  applied  the  double-Gaussian
deconvolution    method    introduced    by   Schneider    \&    Young
(\cite{Schneider})     and    further    elaborated     by    Shafter
(\cite{Shafter}). The  optimal value of the  separation   was determined
from  the diagnostic  diagrams, and  the RV  values measured  for this
Gaussian separation were again  subjected to a power spectrum analysis
in  order  to refine  the  period.  The  photometric  and  RV data  of
SDSS1238 were  analysed for  periodicities using the  Discrete Fourier
Transform (DFT)  code (Deeming \cite{Deeming}) with  a CLEAN procedure
(Roberts et al. \cite{Roberts}).   The H$\alpha$ RV power spectrum is
plotted in the left panel of Fig.\, \ref{fig3}. The maximum peak is at
the 17.88218  day$^{-1}$ frequency,  which corresponds to  the orbital
period  of $  P_\mathrm{orb}=0.05592\pm0.00035$ days. The error of period estimation 
corresponds to FWHI of the main peak in the power spectrum. 
  One day  aliases also
come  up with  lower amplitudes.  The derived  period is  about  4 min
longer  than the 76  min period  recently reported  by Szkody  et al.
(\cite{Szkody2003})  determined  from  only 7  spectra.

Those  RV
measurements were
fitted   with  a   single   sine  curve   with   the  orbital   period
$P_\mathrm{orb}=0.05592$d:
$$       V(t)=\gamma_o+K_1*\mathrm{sin}(2\pi(t-t_0)/P_\mathrm{orb}+\pi),$$       where
$\gamma_o=-3\pm2$km/s is  the systemic  velocity, and $K_1=80\pm4$km/s   the
semi-amplitude  of  the  $H\alpha$  radial  velocity.  
   The  time  of
observation is $t$, and epoch $t_0 =2453112.687\pm0.001$  HJD corresponds
to  the +/- zero  crossing of  the radial  velocity curve.   
Fig.\, \ref{fig3}, right,   shows the radial  velocity  data  of
H$\alpha$ emission line folded  with the spectroscopic orbital period
(0.05592 days) and  its best  sinusoidal fit.  The  central absorption in
H$\alpha$ shows the same RV as the emission line (see
Fig.\, \ref{fig6}, left).

\subsection{Photometric light curves}

We obtained quite  unusual results  from  the  analysis  of the photometric
data presented  in  Fig.\,\ref{fig4}.  \object{SDSS1238}  shows two types
of variabilities:
\begin{quote}
$\bullet$ {\it a long-term  variability} (in the range  of 8-12h)  with $\sim0.45$mag
 amplitude; \\
$\bullet$ {\it  a short-term variability} with half  the orbital period
 ($\sim$40.3 min),  which is  more evident in  the bright part  of the
 long-term variability, reaching $\sim0.35$mag.
\end{quote}
No eclipses are detected in the light curves of the system.

The left  panel of Fig.\,\ref{fig5} shows the  power spectrum computed
for selected  (bright) parts of the  light curve (data  marked by open
circles in  the Fig.\,\ref{fig4}).  Two  practically equal peaks  can be
seen, one at a  frequency that corresponds to 2/P$_\mathrm{orb}$=35.76
day$^{-1}$ and  the other  at 36.76 day$^{-1}$,  being a  1-day alias.
The         main        peak        corresponds         to        half
($P_\mathrm{orb}/2=P^\mathrm{short}_\mathrm{phot}=40.3   \mathrm{min}$)
of the  orbital period.  The  selected photometric data  points folded
with the photometric period ${\rm P^\mathrm{short}_\mathrm{phot}}$ are
shown  in  the  right   panel  of  Fig.\,\ref{fig5}.   The  short-term
photometric   variability  is  sinusoidal   (Fig.\,\ref{fig5},  right)
although  the  amplitude  varies.   Portions  of the  light  curve  of
SDSS1238, together with simultaneous H$\alpha$ emission and absorption
RV measurements and their corresponding fits, are plotted all together
in Fig\,\ref{fig6}.   The short-term, half  orbital period variability
appears to be coherent and  stable with highly variable amplitude. The
amplitude  of  short-term  variability  correlates with  the  cyclical
brightness  variations described here  as long-term  variability.  The
maximums of  the half-orbital period photometric  variability occur at
the  $\phi$=0.375  and  0.875  of  orbital phases  of  the  system  as
demonstrated  in the  right panel  of Fig.\,\ref{fig5}  and  in Fig.\,
\ref{fig6}. We  did not find  any phase shifts from  one observational
run to another (within 2 months).

The  long-term variability  (LTV)  was instantly  detected during  the
first observations  in April 2004  (Table\,\ref{tab1}). The brightness
of  the object  changes up  to 0.4  magnitudes  (Fig.\,\ref{fig4}).  The
period  analysis of  the photometric  data obtained  during  the three
first nights revealed a certain  periodicity of these changes.  In the
May  observations  we  confirmed   the  existence  of  this  long-term
variability; however, the  data from the first two  nights in May were
either out  of phase  or the  periodicity of LTV  was not  strict. The
final two  nights of the  May observations showed good  coherence with
the April data.  The amplitude of LTV in the May observations remained
the same as in April.   The long-term variability was also detected by
Woudt \& Warner (\cite{Woudt})  during several sets of observations in
the  beginning of  2004.  Woudt  \&  Warner (\cite{Woudt})  did not
found any  exact period  of LTV either.   The power spectrum  from the
complete  data sample is  confusing and  difficult to  interpret.  For
natural  reasons observational  runs  are shorter  than any  plausible
period of LTV, and the amount  of observational data is not enough for
any  firm conclusion  about the  existence  of a  strict and  coherent
period.   At the  mean  time it  is  clear that  LTV  has a  cyclical,
quasi-periodic nature at least on short time scales of days about 8-12
hours).  These cycles  are much shorter than the  shortest known dwarf
nova outburst  cycles and neither the  profile of the  light curve nor
the amplitude are similar to the outbursts.

Amazingly, the amplitude of the short-term variability correlates with
this  semi-periodic LTV, and  its amplitude  increases from  less than
$\sim0.1$mag  in  the  faint  part   of  the  light  curve  to  up  to
$\sim0.35$mag in the bright part of the LTV curve.

\subsection{Tomography}

\begin{figure*}[t]
\vspace*{0.2cm}
\setlength{\unitlength}{1mm}
\resizebox{8cm}{!}{
\begin{picture}(100,170)(0,0)
\put (15,85){\includegraphics[width=60mm, clip=]{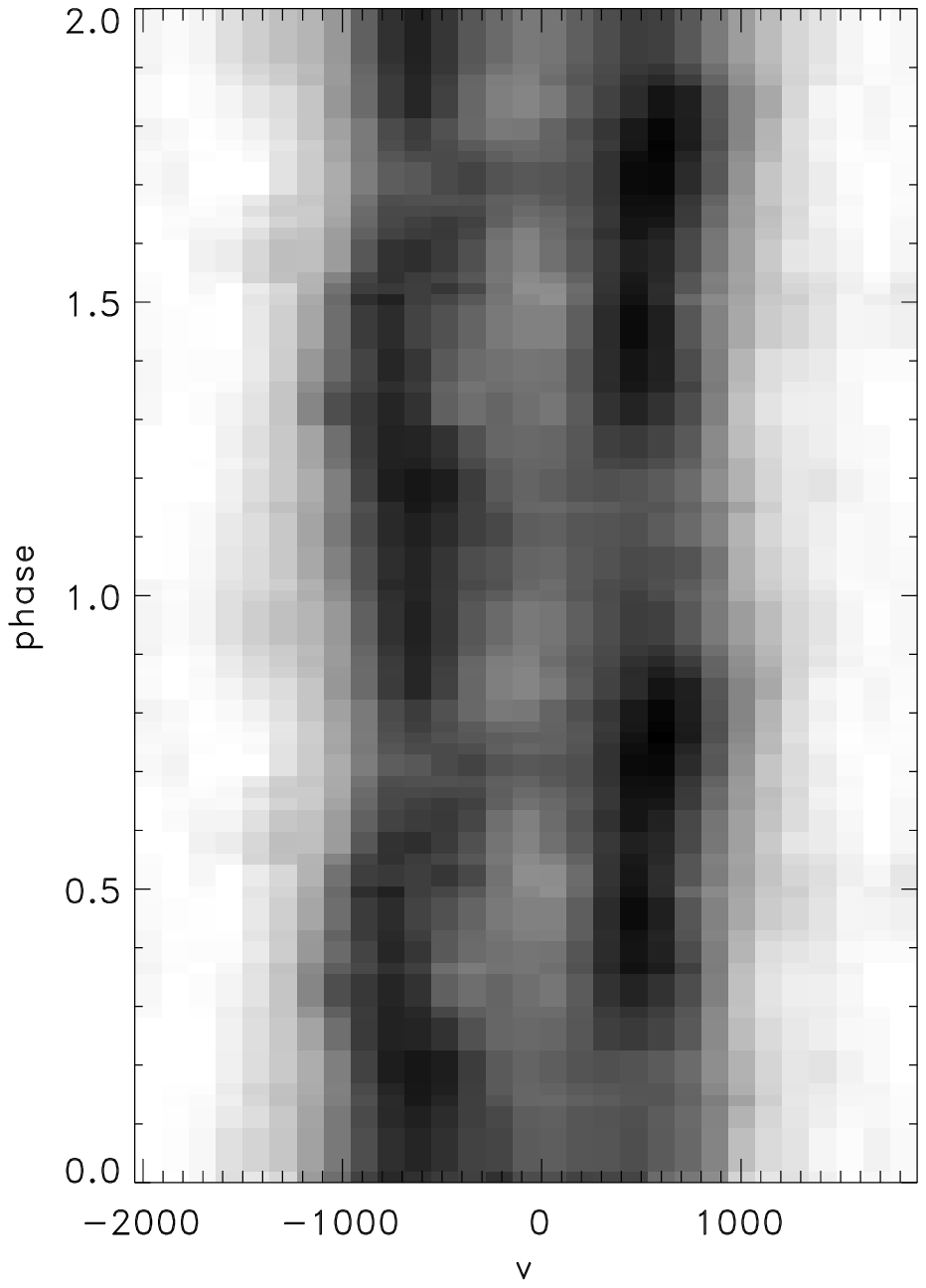}}
\put (80,85){\includegraphics[width=100mm,bb=-100 -28 435 435, clip=]{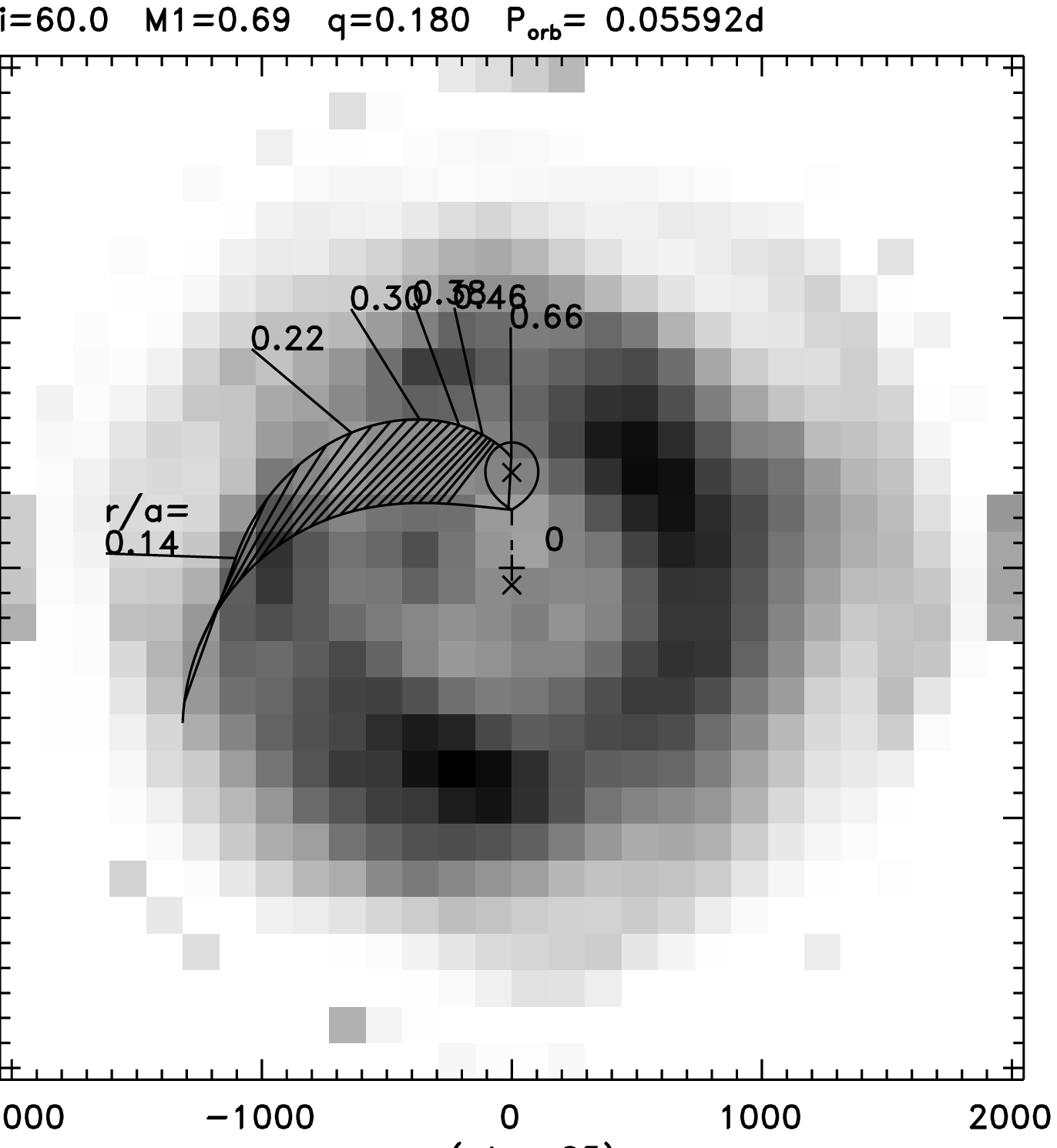}}
\put (15,0){\includegraphics[width=60mm, clip=]{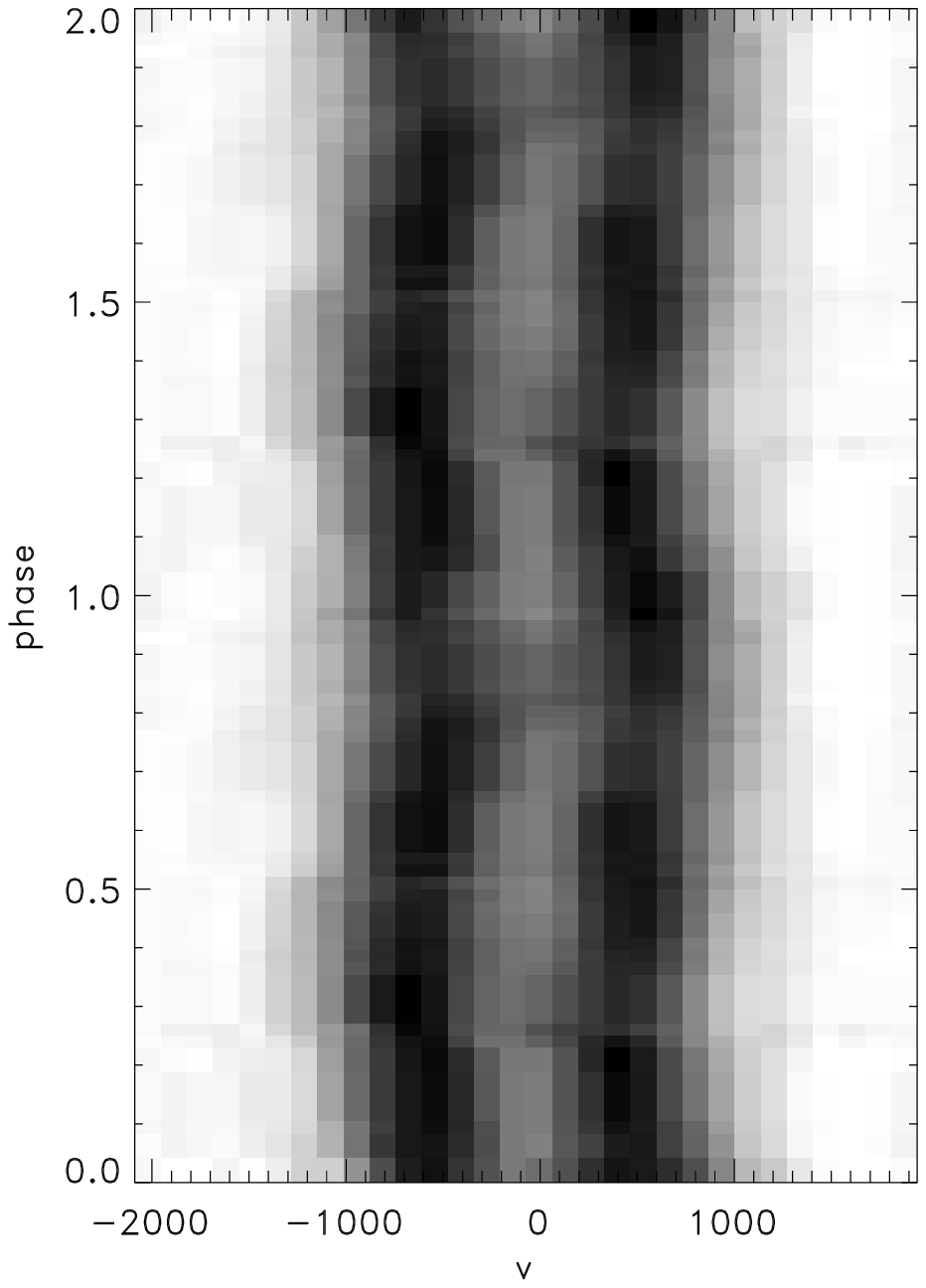}}
\put (80,0){\includegraphics[width=100mm,bb=-100 -28 435 435, clip=]{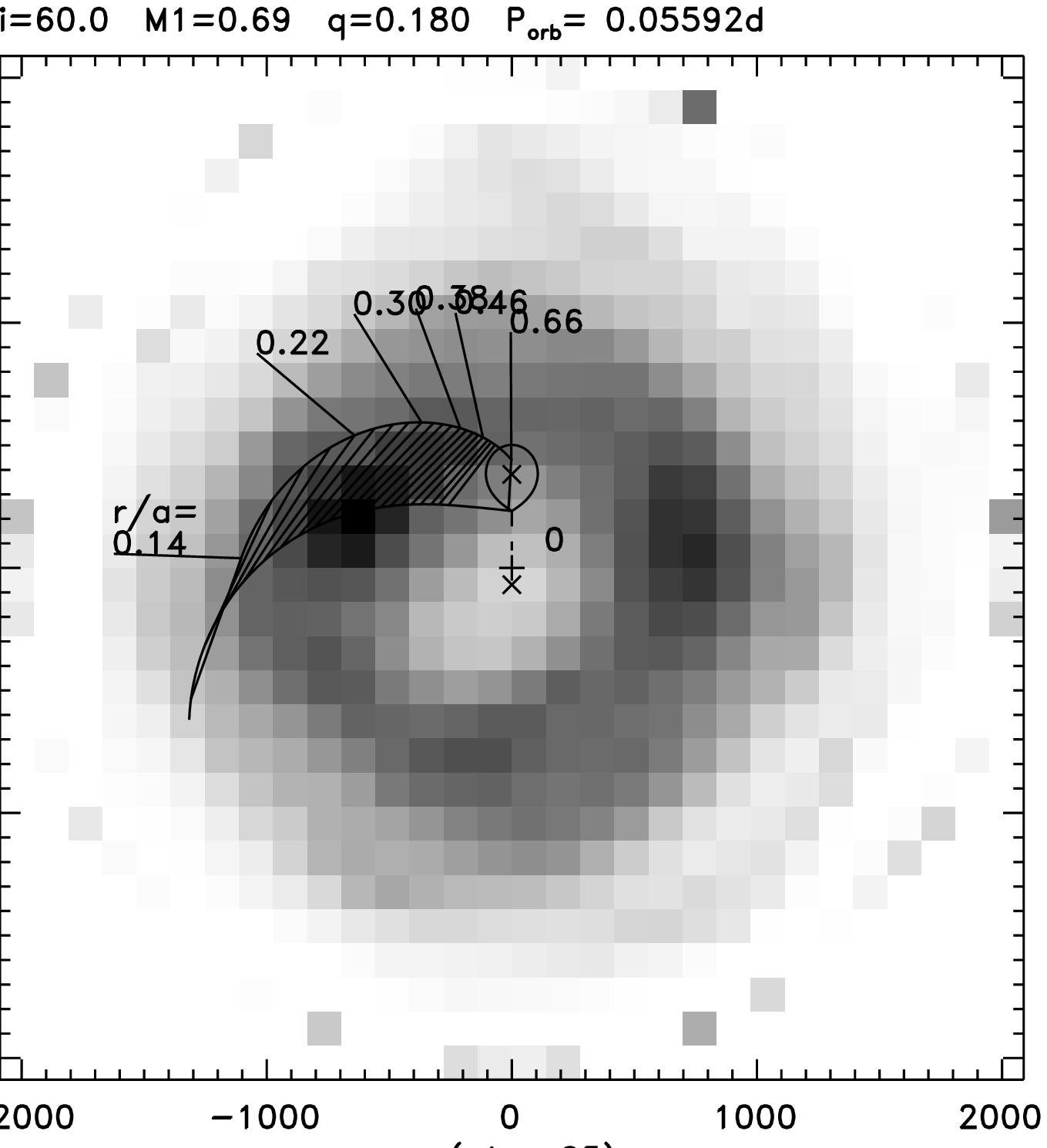}}
\end{picture}}
\caption{The H$\alpha^\mathrm{faint}$, H$\alpha^\mathrm{bright}$ 
 trailed spectra
 folded with the orbital period of the system and their correspondent
 H$\alpha$ Doppler maps (right panels) constructed for the  faint (top) and bright
 (bottom) portions of the long-term variability 
 (see Fig.\, 6).
 Adopted parameters of the system are shown at the top of the Doppler
 maps.}
\label{fig7}
\end{figure*}

The  H$\alpha$  Doppler   maps\footnote{For  details  of  the  Doppler
tomography   method   and    interpretation   see   Marsh   \&   Horne
(\cite{Marsh88}), Marsh (\cite{Marsh02}).   We have generated SDSS1238
Doppler  maps by  using Spruit's  code (Spruit,  \cite{Spruit} Maximum
entropy   method):  http://www.mpa-garching.mpg.de/~henk/pub/dopmap/.}
were  constructed separately  for  the faint  (Fig.\,\ref{fig7},
top panels)  and bright (Fig.\,\ref{fig7},  bottom panels) parts
of  ``LTV-period'' (see Fig.\,\ref{fig6}).  28 spectra  obtained on
April 17  (JD 112) and 15 spectra  obtained on May 19th  (JD 145) were
used for the  faint (20 spectra) and the bright  parts (23 spectra) of
the ``LTV-period''.  The system  parameters used for  Doppler tomogram
mapping    are   shown    on   the    top   of    Doppler    maps   in
Fig.\,\ref{fig7}.    We    used   the    average   mass
$\bar{M}_{wd}=0.69M_\odot$\ for  WDs residing  in CV below  the period
gap (Smith  \& Dhillon  \cite{Smith}).  That leads  to the  mass ratio
$q=M_2/M_1<0.18$ obtained using the Howell et al. (\cite{Howell01})
 relationship
for  secondaries  in  CVs.   The  inclination  angle  is  also  chosen
arbitrarily, as a  compromise for high K$_{em}$ system  in the absence
of eclipses.

 The  Doppler tomograms  show  different states  of  accretion in  the
system.   In  the  low  state the  two  bright symmetric  arc-shaped
regions at  velocity coordinates  (-200km/s, -800km/s) and  (600 km/s,
450km/s)  in Fig.\,\ref{fig7},  (right, top)  look  like the
spiral wave structures in the Doppler maps  of IP { Peg} and { EX} Dra
(Harlaftis    et    al.     \cite{Harlaftis},    Joergens    et    al.
\cite{Joergens}).    In   the   bright   state   (Fig.\,\ref{fig6}   and
\ref{fig7},  right,   bottom),  we  get  only   one  bright  spot
($-600$km/s,  500km/s) at  the expected  place where  the  mass transfer
stream hits the accretion disc.   {This spot moves out from the stream
path if we significantly increase mass of the primary WD, thus proving
that the system parameters are within reasonable values.}  The spot is
completely   absent  in   the  low   state  (Fig.\,\ref{fig7},
right, top).  This difference of the Doppler tomograms built with the
distinct spectra  from the faint and  bright parts of  the light curve
probably testifies  to a rapid  ($\sim$8-12h) change in  $\dot{M}$ and
the accretion disc structure from one to another phase of LTV.

It should  be noted that the  number of spectra,  the phase resolution
(exposure time/orbital  period = 0.12),  and S/N ratio of  spectra are
not high enough to construct reliable Doppler tomograms.  We cannot be
confident of detecting spiral arms, as they might be artifacts arising
from data quality and contrast  of the picture, although brightening of the hot
spot in the bright phase of LTV  is less doubtful. { It appears in the
right place and shape and replicates itself in the $H_\beta$ line (not
shown here).

\section{Discussion}
\label{Discus}

\object{SDSS1238}  is a  remarkable CV  on the  lower edge  of  the CV
period distribution,  which exhibits a number  of interesting features
that define the small class of CVs known as WZ Sge stars.  In addition
to  these  features, it  undergoes  frequent  and cyclical  brightness
changes, defined  here as LTV, that  sets it apart  from other similar
CVs or any other CV as a matter of fact.

The observed period minimum  for CVs with hydrogen-rich secondaries is
about 77min  (Kolb \& Baraffe \cite{Kolb}).  According  to our current
understanding the number of CVs should peak close to the 80 min period
limit,  as they evolve  to the  boundary where  they bounce  back (see
Patterson \cite{Patterson98} and references therein).  If this is true
then only a tiny part  of that presumed stockpile is recovered.  There
are only  about 50  systems known  above the period  minimum to  up to
$\sim$90 min.  Part of the  sample in this period range are \object{SU
Uma} (together with \object{ER Uma})  objects (about 20 of them), some
are magnetic  CVs (10 \object{AM  Her} and 6 \object{DQ  Her} objects)
and  few are firmly  classified as  WZ Sge  stars based  on infrequent
outbursts with distinct high  amplitude and echo re-brightening during
fading from  the outburst (Patterson et  al. \cite{Patterson02}).  The
\object{WZ  Sge} objects  are  probably the  only  ones identified  as
evolved/bounced  systems.    Several  other  stars   are  proposed  as
candidates  to the  \object{WZ Sge}  class based  on  their quiescence
properties.   \object{SDSS1238} bears  certain resemblance  to  WZ Sge
stars, particularly because of absorption features in the spectrum and
the   double  humped   light  curve.    Besides   these  observational
characteristics,  the relatively  low temperature  deduced for the  white dwarf
(see Urban  et al. \cite{Urban}) indicate  that \object{SDSS1238} may
belong to this enigmatic group of objects.

While the reason  for the double-humped light curve  is not understood
well in these systems, it appears  to be more or less widespread among
certain short-period  CVs. True,  in the majority  of them  it appears
only during  the outburst.   In short period  CVs, the  extremely late
type  secondary star  is  so  dim that  it  usually passed  undetected
whether      spectroscopically       or      photometrically.       If
SDSS013701.6-091234.9  may  be  anomalous  in that  sense  (Szkody  et
al.  \cite{Szkody2003},   Pretorius  et  al.    \cite{Pretorius}),  in
\object{SDSS1238} the secondary  is not detected spectroscopically and
the  double  hump  light curve  (detected  in  V  band with  the  same
amplitude as  in the R) can by  no means be caused  by the ellipsoidal
shape  of the  secondary.  As  already mentioned,  in the  majority of
other  systems  the  double-humped  light  curve  is  observed  during
outbursts.  According to  Patterson et  al.   (\cite{Patterson02}) the
double-hump wave  appears in WZ Sge  within 1 day  of outburst maximum
and is then  replaced by a common superhump that  develops in a normal
manner.  The  physical processes  producing double humps  in outbursts
are not clear either.  In Patterson et al.  (\cite{Patterson02}) there
is  a discussion  of possible  causes and  references to  the proposed
models without firm conclusions. Authors favour the model in which the
development of a  strong two spiral-arm structure at  the beginning of
an outburst, when  the 2 : 1 eccentric  resonance is reached, produces
the desired waveform in  the lightcurve (Osaki \& Meyer \cite{Osaki}).
Within  the evidence  supporting  this  model is  the  detection of  a
two-armed spiral  in the Doppler  tomograms of \object{WZ  Sge} during
the first few days of outburst (Steeghs et al. \cite{Steeghs}; Baba et
al. \cite{Baba}).   Another possible explanation  of the double-humped
light  curve  (Patterson  et  al.   \cite{Patterson02})  is  a  sudden
increase in  transfered matter that creates  a hot spot  at the disc's
outer edge.  A hot spot is  a common feature  in CVs and in  many high
inclination systems  it produces a  distinctive shoulder in  the light
curve.  The maximum in the optical light curve origining in such a hot
spot is  around $\phi  = 0.85$, when  the disc  is seen face  on.  The
weaker maximum at opposite $\phi  \sim 0.3$ may be naturally explained
by viewing the  hot spot, partially obstructed by  the accretion disc,
from the opposite  side (Silber et al.  \cite{Silber}).   If we assume
that in low $\dot{M}$ systems in quiescence the disc is optically thin
and tiny, then  the hot spot can be visible through  it and create the
possibility of a double-humped light curve with maxima at phases $\phi
\sim 0.3$ and $\phi \sim 0.8$.

Incidentally   we  detect   both  phenomena:   (i)   an  unhomogeneous
spiral-arm-like structure is  detected at the bottom of  LTV, and (ii)
the  accretion  stream/disc  impact  hot  spot  }  dominating  Doppler
tomogram during the bright part of  LTV.  As in the case of \object{WZ
Sge} (Lasota  et al. \cite{Lasota}; Hameury  et al.  \cite{Hameury}),
the  hot spot model  suggests a  sudden burst  of mass  transfer.  The
authors argue that the irradiation of the secondary provides the basis
for  the  mass transfer  rate  change.   Of  course both  models  were
developed  to explain  the observed  features of  \object{WZ  Sge} and
similar  in outbursts.   However the  phenomenon of  double  humps was
observed in several occasions in the  quiescent state in WZ Sge and AL
Com  (Patterson et  al.   \cite{Patterson96}) and  a  few other  short
period WZ Sge-like systems IY  UMa (Rolfe et al. \cite{Rolfe}), WX Cet
(Rogoziecky      \&      Schwarzenberg-Czerny      \cite{Rogoziecky}),
\object{SDSS013701.06-091234.9}  (Szkody  et  al.   \cite{Szkody2003},
Pretorius  et al.  \cite{Pretorius}), \object{BW  Scl/RX J2353.0-3852}
(Augusteijn      \&     Wisotzki      \cite{Augusteijn};     Abbott et
al. \cite{Abbott}).  In quiescence  there is no  chance that  the disc
extends  as far  as the  2:1 resonance  radius, hence the  bright spot  is a
better  explanation for  the double  humps.  Besides,  in most  of the
other systems, the second hump  is often smaller in amplitude. This is
good evidence  of a bright spot that is  weakened as it is  observed on the
far side of the disc  and through it.  It is convincingly demonstrated
by  Rolfe et  al. (\cite{Rolfe})  using high  S/N spectra  and Doppler
tomography  that the double  hump lightcurve  originates from  the hot
spot.

Our data  on \object{SDSS1238} were obtained certainly  outside of the
outburst regime.   Therefore we simply  assume, by analogy  with other
similar  systems, that  the  double-humped light  curve  profile is  a
result of  increasing hot spot  brightness during the bright  phase of
LTV.  Then,  naturally, the increased  brightness of the spot  and the
system  as  a  whole can  only  be  explained  in terms  of  increased
$\dot{M}$. Why  the mass transfer  rate and hence the  brightness vary
quasi-periodically  on  such  short  time  scales (7-12  hours)  is  a
separate and  very interesting  question. One possible  explanation is
the  irradiation   of  the  secondary   as  favoured  by   Hameury  et
al. (\cite{Hameury}).  Although this  idea was originally suggested to
explain  the triggering of  the superoutburst  in the  \object{WZ Sge}
systems, it  is quite  possible that the  very late type  secondary in
\object{SDSS1238} barely fills its Roche lobe and maintains a delicate
balance.  Then,  the slightest irradiation  from the bright,  hot spot
will expand  the secondary  to fill its  Roche lobe  thereby expelling
increased amounts  of matter toward  the primary.  The  secondary will
then shrink, decreasing the mass  transfer only to be irradiated again
and  to undergo  another cycle  once the  matter reaches  the  edge of
accretion disc and re-brightens it.

In no other system than \object{SDSS1238} is  cyclical brightening
  or LTV detected.  According to our proposed scenario this can be due
  to the  particular mass ratio established  in \object{SDSS1238}. For
  example, in  \object{WZ Sge} the estimated WD  mass is substantially
  higher, which will lead to a much smaller secondary Roche lobe.

\section{Summary}
\label{Summ}
The principal results can be summarised as follows:

1) The  spectrum of  the  system shows  double-peaked Balmer  emission
lines from the  disc and broad Balmer absorption  lines originating in
the  photosphere of  the white  dwarf. The  WD surface  temperature of
15600 K obtained  from the model fit to the spectrum  is within the WD
temperature  range (15\,000  -22\,000K) observed  in the  short period
systems below the period gap (see Urban et al. \cite{Urban})

2) The  spectroscopic  orbital  period  of  the  \object{SDSS1238}  is
0.05592 day+/-0.00035.

3) The system shows two types of photometric variabilities:
\begin{quote}
 $\bullet$  {\it   a  long-term  variability   (LTV)}  with  amplitude
$\approx0.45$  mag and  roughly 8-12  hours  quasi-period;\\ 

$\bullet$ {\it a short-term variability}  with amplitude about $\approx0.35$ mag
in the  bright part of  LTV with a  period two times shorter  than the
orbital period of the system.

\end{quote}
4) The analysis of  Doppler maps constructed for the  bright and faint
parts of LTV  shows that the change of mass transfer  in the system is
the most probable cause of  the long-term variability and that the hot
spot dominates the radiation from the accretion disc at the top of the
LTV.

5) The hot spot is also the probable source of the double humped light
curve. The brightness, as well as  amplitude of the humps, varies in a
cyclical  manner  and is  certainly  a  result  of the  variable  mass
transfer rate.

6) The cause of quasi-periodic flux variability (LTV) and hence of the
variable rate of  mass transfer is not clear, but  could be the result
of the irradiation of the secondary.

\begin{acknowledgements}
This  work  was supported  in  part  by  DGAPA project  IN-110002.  GT
acknowledges the support of  a UC-Mexus fellowship.  R.N. acknowledges
the support  by a PPARC Advanced Fellowship.   VN acknowledges support
of IRCSET  under their basic research  program and the  support of the
HEA-funded  CosmoGrid project.  We  thank anonymous  referee, for  his
comments that lead to an improved presentation of the paper.
\end{acknowledgements}

\end{document}